\documentclass[twocolumn,aps,prx,showpacs,amsmath,amssymb,floatfix,longbibliography,superscriptaddress]{revtex4-2}
\usepackage{graphicx,mathtools,bm,xcolor,array,enumitem,overpic,booktabs,multirow}
\usepackage{natbib}
\usepackage{hyperref}
\hypersetup{
    bookmarks=false, 
    unicode=false, 
    pdftoolbar=false, 
    pdfmenubar=true, 
    pdffitwindow=false, 
    pdfstartview={FitH}, 
    pdftitle={}, 
    pdfauthor={}, 
    pdfsubject={}, 
    pdfcreator={}, 
    pdfproducer={}, 
    pdfkeywords={}, 
    pdfnewwindow=true, 
    colorlinks=true, 
    linkcolor=black, 
    citecolor=black, 
    filecolor=black, 
    urlcolor=black 
}

\newcommand{\argmin}[1]{\mathop{\rm argmin}}

\newcommand{\mytitle}
{Deep learning extraction of band structure parameters from density of states: a case study on trilayer graphene}
\begin{document}
	
\title{\mytitle}
\date\today
\author{Paul Henderson}
\address{School of Computing Science, University of Glasgow, Scotland}
\author{Areg Ghazaryan}
\address{Institute of Science and Technology Austria (ISTA), Am Campus 1, 3400 Klosterneuburg, Austria}
\author{Alexander A. Zibrov}
\address{Department of Physics, Harvard University, Cambridge, MA, USA}
\address{Department of Physics, University of California, Santa Barbara, CA, USA}
\author{Andrea F. Young}
\address{Department of Physics, University of California, Santa Barbara, CA, USA}
\author{Maksym Serbyn}
\address{Institute of Science and Technology Austria (ISTA), Am Campus 1, 3400 Klosterneuburg, Austria}
	
\begin{abstract}
The development of two-dimensional materials has resulted in a diverse range of novel, high-quality compounds with increasing complexity.
 A key requirement for a comprehensive quantitative theory is the accurate determination of these materials' band structure parameters. 
 However, this task is challenging due to the intricate band structures and the indirect nature of experimental probes. 
In this work, we introduce a general framework to derive band structure parameters from experimental data using deep neural networks. We applied our method to the penetration field capacitance measurement of trilayer graphene, an effective probe of its density of states. 
First, we demonstrate that a trained deep network gives accurate predictions for the penetration field capacitance as a function of tight-binding parameters.  Next, we use the fast and accurate predictions from the trained network to automatically determine tight-binding parameters directly from experimental data, with extracted parameters being in a good agreement with values in the literature. We conclude by discussing potential applications of our method to other materials and experimental techniques beyond penetration field capacitance.
\end{abstract}
\maketitle

\section{Introduction}

Electronic band structure of crystalline solids represents a simple yet very rich example of emergence. Under the influence of scattering from the lattice potential, the electron may acquire a different value of effective mass, became massless, and acquire additional quantum numbers such as pseudospin.  In addition, electronic band structure determines basic properties of materials, provided interactions are weak enough~\cite{harrison1989electronic}. Therefore, identifying material parameters that determine the band structure is of crucial importance. From a theoretical point of view {\it ab initio} methods such as density functional theory have achieved enormous success in this direction~\cite{hasnip2014density}. Nevertheless one typically relies on experimental data to quantitatively extract band structure properties. Experimentally there exist numerous ways to access the electronic structure, such as angle resolved photoemission \cite{damascelli2003angle} and X-ray absorption spectroscopy \cite{de2001high}, de Haas-van Alphen effect based on magnetic oscillations \cite{shoenberg2009magnetic}, analyzing reflection and absorption spectra \cite{wooten2013optical}, and electronic transport measurement \cite{mizutani2001introduction}, to name just a few. Despite such a wealth of measurement techniques, matching experimental results with theoretical predictions remains a challenging problem due to the complexity of the band structure, which translates into a large number of involved parameters.

The recent surge of two-dimensional (2D) materials \cite{novoselov20162d} brings 
new aspects 
to the problem of determining band structure. First, often the complexity and the number of parameters in 2D materials is considerably lower compared to their three-dimensional counterparts. Besides, 2D materials feature additional level control such as modifying charge density by gating, and may have an extremely high crystal quality. This opens access to high resolution experimental data, which may potentially be used for precise determination of band structure parameters. A particular example of such data is provided by so-called penetration field capacitance measurements~\cite{eisenstein1992negative}, that effectively probe the density of states (DOS) of the material as a function of carrier density and transverse electric field. Such experimental data has been used to determine material parameters such as hopping matrix elements in several 2D systems \cite{island2019spin,zibrov18}.

Typically, extraction of band structure parameters based on experimental data relies on an efficient solution to what we term the \emph{forward problem}. In the specific example of penetration field capacitance measurements sensitive to the DOS, this means simulating the DOS for specific values of material parameters such as hopping matrix elements entering tight-binding model of the band structure. However, the existence of an efficient solution for this forward problem does not guarantee a fast solution to the \textit{inverse problem}---identification of the physical parameters corresponding to a set of empirical data. The inverse problem is challenging because (i) solving for the best-fitting parameters is a high-dimensional optimization problem
that requires numerous simulations of the forward problem at each step that can quickly become very costly numerically;
and (ii) experimental measurements are typically affected by 
additional factors not easily accounted for in simulation (e.g.~geometric and parasitic capacitance, disorder), meaning that an exact match between the data simulated in the forward problem and that obtained from experiments is not possible.
The typical approach is therefore manual comparison of an experimental dataset with a large number of simulated ones, relying on physical intuition of which features are important. This process is laborious and computationally expensive~\footnote{This approach is aptly referred to as \textit{graduate-student descent}.}, calling for the development of more efficient and systematic approaches.

In this work we present a machine learning based method that automates the process of comparing numerical simulation and experimental data, so the physical parameters of the band structure that gave rise to a particular experimental dataset can be determined with minimal human effort. Recently machine learning and artificial neural network techniques have seen various applications in the realm of physical sciences~\cite{Carleo2019machine}. In condensed matter physics, artificial neural networks have been used to represent quantum states~\cite{Carleo2017solving,Choo2018symmetries} and learn these states from available data~\cite{cai2018approximating,borin2020approximating}. In a different direction, recently machine learning models were use for photonic crystals band diagram prediction and gap optimisation~\cite{Christensen:2020aa,Itin22,Valag20-2}. Despite a large number of more theoretical applications, machine learning approaches are only starting to be employed in analysis of experimental data.  Recent examples include identification of quantum phase transitions~\cite{rem2019identifying} and hidden orders from experimental images~\cite{zhang2019machine}. These few examples highlight the strong potential of machine learning based approaches on experimental data, that we further exploit in the present work.

\begin{figure*}
	\includegraphics[width=\linewidth]{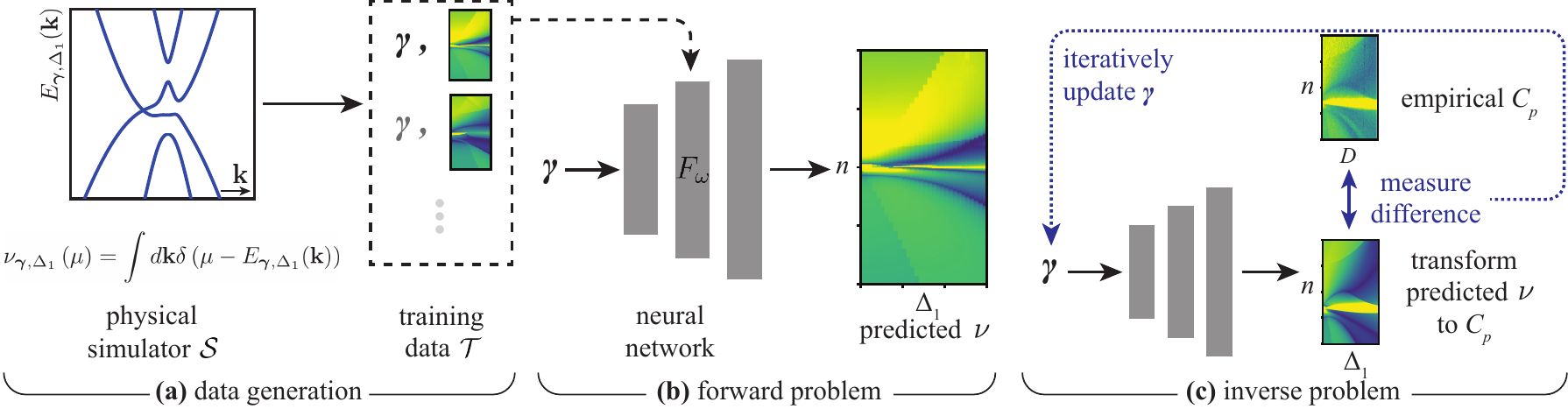}
	\caption{
		Our approach to the direct extraction of the band structure parameters uses datasets obtained with a physical simulator shown in panel (a) to train a DNN in panel (b), thus providing a more efficient solution to the forward problem. The DNN-based simulator is used for a gradient optimization of the band structure parameters, allowing to extract their values from experimental data in panel (c).  
	}
	\label{fig:outline}
\end{figure*}

A conceptual overview of our approach is shown in Fig.~\ref{fig:outline}.
To extract the band structure parameters from experimental data, we first train a deep neural network (DNN)~\cite{Goodfellow2016book,Bishop2006pattern} that solves the forward problem by replicating the numerical calculation of the DOS (Section~\ref{sec:method-fwd}).
To this end we use the simulation of the experimental data shown in Fig.~\ref{fig:outline}(a). In the particular example of penetration field capacitance data considered here, the simulator uses the band structure parameters, the asymmetry potential between two edges of the system (physically equivalent to transverse electric field) and the chemical potential as input parameters. As an output we get charge density and from that determine the DOS by differentiating density with respect to chemical potential.
A set of simulated data is used to train the DNN in Fig.~\ref{fig:outline}(b). Constructed in a way to efficiently replace the data simulator, the DNN acts as a function that takes the band structure parameters, the asymmetry potential and directly charge density as input, and outputs the corresponding DOS. It is constructed by learning from a large dataset of simulation results, optimising its output to always match that of the simulator.  The resulting DNN represents a fast and differentiable replacement for the physical simulation. It can therefore be used to efficiently solve the inverse problem (Section~\ref{sec:method-inverse}). In particular, the values of parameters that gave rise to a given dataset are extracted using gradient-based optimisation in Fig.~\ref{fig:outline}(c), where we iteratively modify the band structure parameters until the DNN's output matches the provided DOS values.

The task of mapping a vector of inputs (e.g.~band structure parameters) to a continuous output (e.g.~DOS) is known as \textit{regression} in machine learning \cite{Bishop2006pattern}.
A DNN implements such a mapping as a series of chained matrix multiplications (`layers') interleaved with elementwise non-linear functions (`activations').
Each layer multiplies the vector of outputs from the previous by some weight matrix, to give an updated vector~\cite{Goodfellow2016book}; the final layer typical yields a single value.
The weight matrices are optimized (`trained') using first-order optimization (e.g.~gradient descent), such that the overall mapping from inputs to output approximates some function defined by a \textit{training dataset} of inputs and desired outputs.
The celebrated \textit{universal approximation theorem}~\cite{hornik89universal} proves that a neural network with just two layers (but unbounded width) can represent any smooth function.
More recently, it has been shown that this is also true for a neural network of bounded width (but unbounded depth)~\cite{kidger20universal}.
In practice, even finite-sized DNNs have proven very successful in approximating complex functions in many domains of science. 
One of our contributions is to show a DNN can also provide an accurate estimate of DOS given band-structure parameters, field strength, and chemical potential.

Since our final goal is to determine the band-structure parameters from experimental measurements of penetration capacitance, it might seem natural to train the DNN for exactly this task (the inverse problem), instead of the forward problem.
However, this is not possible in practice.
We have only a \emph{single} experimental dataset, for which the parameters are unknown, whereas machine learning techniques require a large dataset of training examples (with the true output known) to learn the desired mapping from.
If instead we trained on easy-to-obtain simulated data, the resulting model would not work on experimental data
since the latter is significantly different from the former both in terms of the relative magnitude of features in the data,
and the locations of features such as edges. These differences may arise since the simulator uses a simplified effective model of the material and does not account for screening at the microscopic level, disorder, strain, experimental uncertainties, and possibly other ingredients.
In contrast, statistical learning theory requires that the training and test data be drawn from the same distributions if a trained model is to work on the latter~\cite{vapnik91principles}.

As a specific application, we demonstrate the framework outlined above on Bernal stacked ($ABA$) trilayer graphene. For this material both band structure parametrization \cite{dresselhaus1981intercalation,serbyn2013new} and high quality experimental measurements are readily available \cite{zibrov18}, calling for an accurate extraction of the band structure parameters.  The determination of the band structure was performed by tour de force manual fitting in an earlier work~\cite{zibrov18}, thus allowing us to benchmark our approach.  
First, we train the DNN and show it gives an efficient and accurate surrogate for numerical calculation of the DOS for this system, for a wide range of band structure parameters (Section~\ref{sec:app-fwd}). Next, we use the DNN for automatically solving the inverse problem of determining the physical parameters giving rise to certain values of the penetration capacitance (Section~\ref{sec:app-inverse}). Finally, we apply this  to experimental data from Ref.~\cite{zibrov18} by exploiting techniques from computer vision, that allow matching important features of the measurements (e.g.~Van Hove singularity peaks and jumps of DOS), while ignoring features that differ between experimental and simulated data (e.g.~measurement noise and discreteness of calculation grid) (Section~\ref{sec:app-expt}). The resulting values of parameters agree within error bars with the estimates from the literature, thus providing a particular benchmark for our approach.

The paper is organized as follows: Section~\ref{sec:method} describes the general structure of the DNN for the forward problem and our minimization approach for inverse problem. Section~\ref{sec:app} applies the framework constructed in Section~\ref{sec:method} to $ABA$ graphene simulation results for the forward and inverse problems, eventually utilizing it for extracting band parameters from experimental results. Finally Section~\ref{sec:discussion} is devoted to discussion of the main results and generalization of the model to other systems.

\section{Method}
\label{sec:method}

We assume access to a simulator $\mathcal{S}$ (see Fig.~\ref{fig:outline}) that in the particular example of the penetration field capacitance calculates the charge density $n$ and density of states $\nu=\frac{\partial n}{\partial \mu}$ given band structure parameters $\bm\gamma$, interlayer asymmetry $\Delta_1$, and chemical potential $\mu$.
We assume $\bm\gamma \in \Gamma$, where $\Gamma$ defines a physically-plausible range for those parameters, and similarly that $(\Delta_1 ,\, n) \in P$. The approach presented here is general, while the specific physical meaning of these parameters will be discussed in Section~\ref{sec:app}. Typically the simulator will be slow to evaluate, making it difficult to use `in the loop' for solving the inverse problem, of finding the physical parameters $\bm\gamma$ corresponding to observed data.
We shall instead use $\mathcal{S}$ to generate training data represented by tuples $(\bm\gamma ,\, \Delta_1 ,\, n ,\, \nu)$ for a machine learning regression model---a deep neural network---$F_\omega$, that will be trained to approximate $\mathcal{S}$ in Section~\ref{sec:method-fwd}.
We shall then use the resulting DNN $F_\omega$ when solving the inverse problem in Section~\ref{sec:method-inverse}.

\subsection{Forward problem}
\label{sec:method-fwd}

We introduce a function
	$F_\omega$ that maps $\bm\gamma$, $\Delta_1$ and $n$ to the DOS.
We choose $F_\omega$ to be a deep neural network (DNN) \cite{Goodfellow2016book}, with weights $\omega$; these weights are free parameters that determine the function it represents.
Our goal is that $F_\omega$ matches $\mathcal{S}$ as closely as possible for all relevant values of $\bm\gamma$, $\Delta_1$ and $\mu$,
i.e.~if the simulator returns $n$ and $\nu$ for given $(\bm\gamma ,\, \Delta_1 ,\, \mu)$ and if $(\Delta_1 , n) \in P$ are within the domain of physically realistic parameters, then DNN approximates well the DOS, $F_\omega(\bm\gamma , \Delta_1 , n) \approx \nu$.
The network weights $\omega$ would ideally be set to minimise the absolute difference between the network's predicted values and those $\nu$ from the simulator, over the entire parameter space $\Gamma \times P$:
\begin{equation}\nonumber
	\omega = \argmin_{\omega'}
	\int_{ (\Delta_1 ,\, n) \in P, \bm\gamma \in \Gamma }
	\big| F_{\omega'}( \bm\gamma ,\, \Delta_1 ,\, n ) - \\
	\nu \big|
	\, \mathrm{d} \bm\gamma \, \mathrm{d} \Delta_1 \mathrm{d} n
\end{equation}
In practice we instead minimise the mean error over a finite \textit{training set} \cite{Bishop2006pattern} of points $\mathcal{T} \subset \Gamma \times P$ at which we have precomputed $n$ and $\nu$ using the simulator $\mathcal{S}$, i.e.
\begin{equation}
	\label{eq:fwd-min}
	\omega = \argmin_{\omega'} \;
	\frac{1}{| \mathcal{T} |}
	\sum_{ ( \bm\gamma ,\, \Delta_1 ,\, n ,\, \nu ) \in \mathcal{T} }
	\big| F_{\omega'}( \bm\gamma ,\, \Delta_1 ,\, n) - \nu \big|.
\end{equation}
To solve this optimisation problem,
the weights $\omega$ are initialised using the heuristic of Ref.~\cite{glorot10understanding}, then iteratively updated using the first-order stochastic-gradient optimiser Adam \cite{kingma15adam} with a minibatch size of 512 and learning rate (step size) of $10^{-3}$.
We use a DNN with five fully-connected layers of 512 units each, with ELU nonlinearities \cite{clevert15elu}, layer normalisation \cite{ba16layernorm}, and residual connections \cite{he16resnet}.
For the input layer, we use Fourier feature embedding with four octaves \cite{tancik20fourier}; for the output layer, we use a single linear unit, see Appendix~\ref{sec:architecture} for details.
We select these architectural parameters, and determine when to stop training the DNN, based on its performance on a separate \textit{validation set}, that is disjoint from $\mathcal{T}$.
We implemented the DNN using the TensorFlow library \cite{abadi15tensorflow}.
Codes for generating data, training, and evaluating the DNN are publicly available~\cite{GITURL}.

\subsection{Inverse problem}
\label{sec:method-inverse}
Suppose we have an experimental dataset that provides measurements of penetration capacitance $C_p(D ,\, n)$, which is a quantity that is sensitive to the DOS, as we discuss below. The measurements are acquired while varying $(D ,\, n)$ over some finite set $Q$;
here $D$ is the strength of an externally-applied electric field that affects $\Delta_1$.
The inverse problem is to determine the band structure parameters $\bm\gamma$ for the system, i.e.~to find the setting for $\bm\gamma$ for which the simulated results are the closest to the experimental ones.
We use the trained DNN $F_\omega$ in place of the simulator $\mathcal{S}$; we therefore seek
\begin{equation}
	\label{eq:inv-min}
	\bm\gamma = 
	\mathop{\rm argmin}_{ \bm \gamma' \in \Gamma} \,
	\min_{\bm\alpha ,\, \bm\beta} \,
	\frac{1}{|Q|}
	\sum_{(D, n) \in Q}
	d \! \left( C_p^*(D , n) ,\, C_p(D ,n) \right)
\end{equation}
where $d$ is a metric quantifying the difference between experimental and predicted values, and we introduced a function $C_p^*(D ,n)$ that relies on the electrostatic model to convert the DOS approximated by the DNN into the penetration field capacitance value,
\begin{equation}
	\label{eq:cp-from-nn}
C_p^*(D , n) = \frac{\beta_1}{\beta_2 + F_\omega(\bm\gamma' ,\, \alpha_1 D + \alpha_3 D^3 , n )},
\end{equation}
with $\alpha_{1,3}$ and $\beta_{1,2}$ being parameters that encode the screening and electrostatic characteristics of the experimental setup respectively.
This function and intuition behind these additional parameters is described in detail in Section~\ref{sec:app-physics}.

We solve the minimization problem (\ref{eq:inv-min}) using the Adam optimiser \cite{kingma15adam} \footnote{We also experimented with the constrained trust-region method of Ref.~\cite{byrd99interior}, and the quasi-Newton method L-BFGS-B~\cite{zhu97lbfgs}, both as implemented in \texttt{scipy} \cite{virtanen20scipy}. However, noise in the experimental data caused both these methods to perform poorly, becoming trapped in local optima.}.
This optimiser makes use of the gradient of the objective with respect to $\bm\gamma$; to calculate this, we use reverse-mode automatic differentiation on the objective in Eq.~(\ref{eq:inv-min}), similar to the back-propagation process used when training the DNN \cite{rumelhart86backprop}, but now with the weights $\omega$ held fixed and gradients instead propagated to the inputs~$\bm\gamma$.

\section{Application to ABA graphene}
\label{sec:app}
In this section we discuss a specific application of our method---determining the band structure of Bernal stacked trilayer (``$ABA$'') graphene.
We first describe the physical system and how it is simulated in Section~\ref{sec:app-physics}.
Then, we discuss the dataset that we generate using this simulator (Section~\ref{sec:app-data}) and use to train the DNN.
Finally, we discuss the performance of the trained DNN when solving the forward (Section~\ref{sec:app-fwd}) and inverse (Section~\ref{sec:app-inverse}) problems, and how we apply it to experimental data (Section~\ref{sec:app-expt}).

\subsection{Physical system and simulation}
\label{sec:app-physics}
 The band structure of $ABA$ graphene can be decomposed through rotation of the basis into monolayer and bilayer graphene type sectors, which get coupled through a displacement field applied between the layers. The Hamiltonian matrix takes the form \cite{serbyn2013new}
\begin{equation}
H_\mathbf{k}=\left(\begin{array}{cc}
H_\mathrm{SLG} &V \\
V^T & H_\mathrm{BLG}
\end{array}\right),
\label{HamABA}
\end{equation}
where
\begin{eqnarray}\nonumber
H_\mathrm{SLG}&=\left(\begin{array}{cc}
\Delta_2-\frac{\gamma_2}{2} & v_0\pi^\dagger \\
v_0\pi & -\frac{\gamma_5}{2}+\delta+\Delta_2
\end{array}\right),
\\ \nonumber
H_\mathrm{BLG}&=\left(\begin{array}{cccc}
\frac{\gamma_2}{2}+\Delta_2 & \sqrt{2}v_3\pi & -\sqrt{2}v_4\pi^\dagger & v_0\pi^\dagger \\
\sqrt{2}v_3\pi^\dagger & -2\Delta_2 &v_0\pi &-\sqrt{2}v_4\pi \\
-\sqrt{2}v_4\pi & v_0\pi^\dagger &\delta-2\Delta_2 & \sqrt{2}\gamma_1 \\
v_0\pi & -\sqrt{2}v_4\pi^\dagger & \sqrt{2}\gamma_1 & \frac{\gamma_5}{2}+\delta+\Delta_2
\end{array}\right), 
\\ \nonumber
V&=\left(\begin{array}{cccc}
	\Delta_1 &0 &0 &0 \\
	0 &0 &0 &\Delta_1
\end{array}\right).
\end{eqnarray}
Here $v_i=\sqrt{3}a\gamma_i/2$ and $a=2.46\,\AA$ is the lattice constant. $\gamma_i$ and $\delta$ are the hopping and on-site potential parameters of the physical system. $\pi=\xi k_x+ik_y$ is the crystal imaginary momentum measured with respect $K^{\pm}$ valley points labeled by $\xi=\pm1$. $\Delta_1$ describes the effect of the external field and $\Delta_2$ charge asymmetry between external and internal layers.

For producing simulated data we calculate density and DOS on a grid using eigenvalues of Hamiltonian (\ref{HamABA}) $\epsilon_\mathbf{k}$ for each value of chemical potential $\mu$
\begin{eqnarray} \label{Eq:density}
n\left(\mu\right)&=4\frac{S_k}{\left(2\pi\right)^2}\frac{1}{N}\sum_{\mathbf{k}}n_F\left(\epsilon_\mathbf{k}-\mu\right), \\
\label{Eq:DOS}
\nu\left(\mu\right)&=4\frac{S_k}{\left(2\pi\right)^2}\frac{1}{N}\sum_{\mathbf{k}}n^\prime_F\left(\epsilon_\mathbf{k}-\mu\right),
\end{eqnarray}
where $4$ accounts for spin and valley degeneracy, $S_k$ is the area of the Brillouin zone sampled in the grid, $N$ is the number of grid points and $n_F\left(x\right)=1/\left(e^{x/T}+1\right)$ is the Fermi-Dirac distribution. We use finite temperature $T=0.025\,\mathrm{meV}$ in the calculation to smoothen singularities of DOS at Van Hove singularity points, and a hexagonal grid with a momentum cutoff $ka=0.15$ and $7.5\times10^5$ grid points.

To obtain the dependence of the experimentally measured penetration field capacitance on the DOS we can use the following formula \cite{hunt2017direct,zibrov2017tunable,zibrov18}:
\begin{equation}
C_p=\frac{c_tc_b}{c_t+c_b+e^2\nu},
\end{equation}
where $c_b$ and $c_t$ is the geometric capacitance of bottom and top gate, $e$ is the charge of the electron.
However, this is a simplified formula which ignores layer polarization change due to the applied electric field $D$, and views the trilayer graphene as a single layer system. Therefore, to allow a more general relationship, we introduced the parameters $\beta_1$ and $\beta_2$ in (\ref{eq:cp-from-nn}). Besides linear screening, we allow for non-linear component, so 
\begin{equation}\label{Eq:screen}
\Delta_1=\alpha_1D+\alpha_3D^3.
\end{equation} 
Due to the presence of a parasitic capacitance, we model the relation between experimentally measured and calculated charge density by
\begin{equation}
n^\prime=n+\left(\mu-\mu\left(\Delta_1=0,n=0\right)\right)/\eta,
\label{eq:DensConv}
\end{equation}
where $\eta$ is the inverse of parasitic capacitance.

In the figures, simulation results are presented in terms of inverse of DOS $\nu^{-1}$ in units of $\mathrm{eV}\cdot A_\mathrm{u.c}$, where $A_\mathrm{u.c}=\sqrt{3}a^2/2$ is the area of graphene unit cell. Charge density is always presented in units of $\mathrm{cm^{-2}}$. Band structure parameters $\gamma_i$ and parameters $\delta$, $\Delta_1$, $\Delta_2$ and $\mu$ are presented in $\mathrm{meV}$.

\begin{figure*}[t]
	\includegraphics[width=0.94\linewidth]{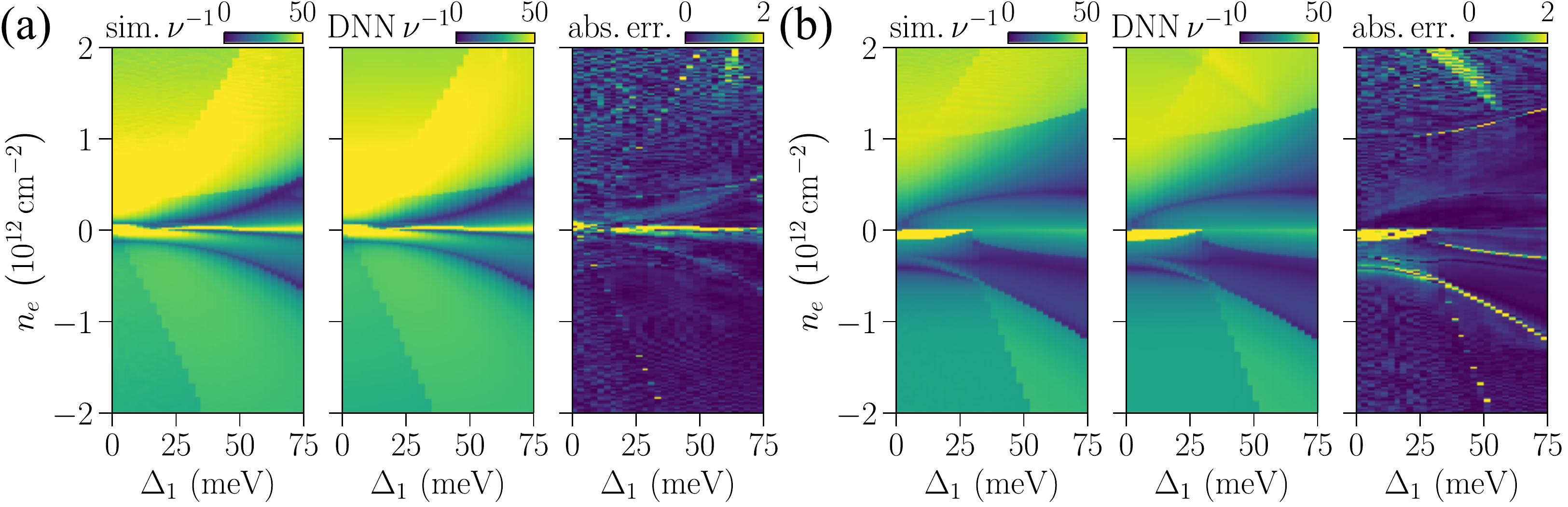}
	\caption{
	Predictions of the inverse DOS, $\nu^{-1}$, by our DNN, compared with results from a numerical simulator.
	(a) and (b) corresponds to the band-structure parameters $\gamma_2=-4.3$, $\gamma_3=-342$, $\gamma_4=133.4$, $\Delta_2=-0.4$ and $\gamma_2=-15.6$, $\gamma_3=-248.8$, $\gamma_4=104.9$, $\Delta_2=11.2$ respectively. Both of the sets were not present in the training data for the DNN.
	For each we show:
	(left) $\nu^{-1}$ as a function of field strength $\Delta_1$ and electron charge density $n$, as predicted by a simulator;
	(middle) $\nu^{-1}$ as predicted by our DNN; ideally this would exactly match the simulator output;
	(right) the absolute difference between the simulator and DNN values.
	We see good correspondence between the DNN's outputs and the simulator, indicating that it has been successfully trained to replicate the simulator output, at much lower computational cost.
	}
	\label{fig:forward}
\end{figure*}

\subsection{Data generation}
\label{sec:app-data}

\begin{table}[b]
	\centering
	\caption{Range of physical parameter and grid step used in the calculations.}
	\label{ParTab}
	\begin{tabular}{p{0.22\columnwidth}p{0.22\columnwidth}p{0.22\columnwidth}p{0.22\columnwidth}}
	\hline\hline
	Parameter & Minimum (meV) & Maximum (meV) & Step (meV) \\
	\hline
	$\gamma_0$ & $3100$ & $3100$ & -- \\
	$\gamma_1$ & $380$ & $380$ & -- \\
	$\gamma_2$ & $-25$ & $-1$ & $3$ \\
	$\gamma_3$ & $-340$ & $-242$ & $7$ \\
	$\gamma_4$ & $100$ & $177$ & $7$ \\
	$\gamma_5$ & 50 & 50 & -- \\
	$\delta$ & 35.5 & 35.5 & -- \\
	$\Delta_2$ & $0$ & $11$ & $1$ \\
	\hline\hline
	\end{tabular}
\end{table}
Using the simulation $\mathcal{S}$ described above, we generate data to train and evaluate the DNN.
In order to define a suitable grid, we choose the space $\Gamma$ of valid physical parameters using physical insights into the meaning of the tight-binding parameters.
Since $\delta$ and $\gamma_5$ introduce only a simple energy shift on the spectrum for a considerable range of values, we fix them to $\delta=35.5\,\mathrm{meV}$ and $\gamma_5=50\,\mathrm{meV}$. We also fix $\gamma_0=3100\,\mathrm{meV}$ and $\gamma_1=380\,\mathrm{meV}$; these are intralayer and direct interlayer hopping matrix elements, which were studied extensively in graphite, and in monolayer and bilayer graphene.
The remaining tuning parameters that constitute $\bm\gamma$ are $\gamma_2$, $\gamma_3$, $\gamma_4$, $\Delta_2$ and $\eta$.
These parameters are allowed to take values in the ranges shown in Table~\ref{ParTab}.
Finally, we define $P$, the region of acceptable parameters $\Delta_1$ and $n$, according to $0 \leq \Delta_1 \leq 75 \, \mathrm{meV}$ and $|n| < 4 \times10^{12}\,\mathrm{cm^{-2}}$, corresponding to the physically achievable range of transverse electric fields and densities respectively.

To generate  datasets for training, validation, and testing of the DNN, we iterate over a grid of $\gamma_2$, $\gamma_3$, $\gamma_4$ and $\Delta_2$, with the respective increments for each specified in the last column of Table~\ref{ParTab}.
The increment is chosen so that for each parameter the grid contains on the order of 10 points.
We also add an additional random offset of up to half the step size to each value, to avoid the DNN overfitting on points of the grid.
We then partition this set of parameter vectors into three disjoint subsets:
we randomly assign 75\% of parameter values to the training set $\mathcal{T}$, 10\% for validation and the remainder for testing~\footnote{Note that the validation set is used for determining convergence and selecting the network architecture, while all results are reported on the testing set; this avoids overfitting to the testing data.}.

For each value of the structure parameters, we vary $\Delta_1$ and $\mu$ in the intervals $\left[0,75\right]\,\mathrm{meV}$ and $\left[-115,115\right]\,\mathrm{meV}$ with steps $2.5\,\mathrm{meV}$ and $0.025\,\mathrm{meV}$ respectively.
We calculate the corresponding values of $n$ and $\nu$ using $\mathcal{S}$ as specified in Eqs.~(\ref{Eq:density})-(\ref{Eq:DOS}).
Since the charge density $n$ for different values of the parasitic capacitance may be determined straightforwardly from that when it is zero, we fix the inverse parasitic capacitance $\eta = \infty$.
This yields a dataset of tuples of values $(\gamma_2 ,\, \gamma_3 ,\, \gamma_4 ,\, \Delta_2 ,\, \Delta_1 ,\, n ,\, \nu)$.
We discard tuples where $n \notin P$ (which is beyond our range of interest), or both $|n| < 0.01\times10^{12}\,\mathrm{cm^{-2}}$ and $\nu < 0.01\mathrm{eV}\cdot A_\mathrm{u.c}$ (i.e.~very close to the charge neutrality point and the gapped region, respectively).
During the training process itself, we sample the value of inverse parasitic capacitance, $\eta$, from a log-uniform distribution on the interval $\left[50,5000\right]\,\mathrm{10^{-12} \, meV\times cm^2}$, independently for each element of each minibatch, and transform $n$ according to (\ref{eq:DensConv}).
In total our training set contains $1.38 \times 10^9$ points from 14580 settings of the band structure parameters; there are $184 \times 10^6$ points in the validation set, and $276 \times 10^6$ in the test set.

\subsection{Forward problem: calculating $\nu$}
\label{sec:app-fwd}
Using the dataset of simulation results for $ABA$ graphene described above, we train a DNN following the protocol in Section~\ref{sec:method-fwd}, to predict the density of states~$\nu$.

To evaluate the accuracy of the trained DNN, we use it to predict $\nu$
for points in the held-out test set,
i.e.~for parameters $(\gamma_2, \gamma_3, \gamma_4, \Delta_2) \notin \mathcal{T}$ that were not contained in the training dataset.
Quantitatively, our method achieves a mean absolute error in $\nu$ of $0.00288\,\mathrm{eV}\cdot A_\mathrm{u.c}$, corresponding to a mean relative error of $5.95\%$.
We further analyzed the cause of these errors, by resimulating a random subset 300 of plots in the test set using a finer momentum grid with $6.75\times10^6$ points and cutoff $ka=0.2$.
Comparing the DNN's predictions to these more accurate simulations, the absolute error reduces to $0.00067\,\mathrm{eV}\cdot A_\mathrm{u.c}$, corresponding to a mean relative error of $1.32\%$.
We also measure the relative error of the lower-resolution simulations forming our dataset, with respect to the 300 higher-resolution plots, and find this is $5.97\%$.
Remarkably, the DNN's predictions are therefore slightly more accurate than the lower-resolution simulations used for its training data.
We hypothesize that this is due to the DNN's inductive bias towards learning smooth functions discouraging it from learning the very variable high-frequency artifacts that arise due to the coarser momentum grid. Therefore, in the following figures we show simulator results for higher resolution momentum grid, although DNN training was performed using lower-resolution data.

Examples of outputs from the DNN, alongside corresponding outputs from the simulator, are shown in Figure~\ref{fig:forward}.
We also visualize the mean absolute difference between the DNN's predictions, and those of the simulator.
We see that
the DNN
produces outputs closely matching the simulator, most importantly accurately incorporating features such as discontinuities and (smoothened) singularities that originate from the changes of Fermi surface topology upon tuning density, $n$ or $\Delta_1$. For instance, the diagonal feature for negative densities in Fig.~\ref{fig:forward}(a) corresponds to the disappearance of the hole pocket upon increasing density, while parabola-shaped blue line in the negative density region corresponds to the Van Hove singularity where DOS would show a logarithmic divergence in absence of cutoffs introduced due to finite temperature and finite grid size in momentum space.
The plots of absolute differences show that the errors are typically very small in regions of uniform $\nu^{-1}$, and are dominated by small misalignments of the features that correspond to DOS discontinuities or singularities.

We also measure the difference in computation time for the simulator and the DNN.
Running with four threads on a 3.6GHz Intel Core i7-9700K CPU, the simulator takes 1445\,s 
to evaluate $\nu$ for the grid of $(\Delta_1, \mu)$ specified in Section~\ref{sec:app-data} and the fine momentum grid.
Running the DNN on the same CPU with four threads takes just 1.1\,s.
Moreover, it does so with higher-resolution sampling of the range of values of $n$ that are of interest, due to taking $n$ as an input instead of $\mu$, hence avoiding the need to discard calculations where $n$ is outside the range of interest.

\begin{figure*}[t]
	\hspace{0.5cm}
	\begin{tabular}{lcccc}
		& $\gamma_2$ & $\gamma_3$ & $\gamma_4$ & $\Delta_2$ \\
		\hline
		true &$ -15.58 $&$ -314.1$ & 161.1 & $-0.15$ \\
		predicted & $-15.36$ & $-313.3$ & 160.2 & $-0.16 $\\
	\end{tabular}
	\begin{tabular}{lcccc}
		& $\gamma_2$ & $\gamma_3$ & $\gamma_4$ & $\Delta_2$ \\
		\hline
		true & $-18.64$ & $-265.7$ & 114.8 & 5.30 \\
		predicted &$ -18.56$ & $-265.0$ & 115.1 & 5.28 \\
	\end{tabular}
	\\
	\includegraphics[width=0.5\linewidth]{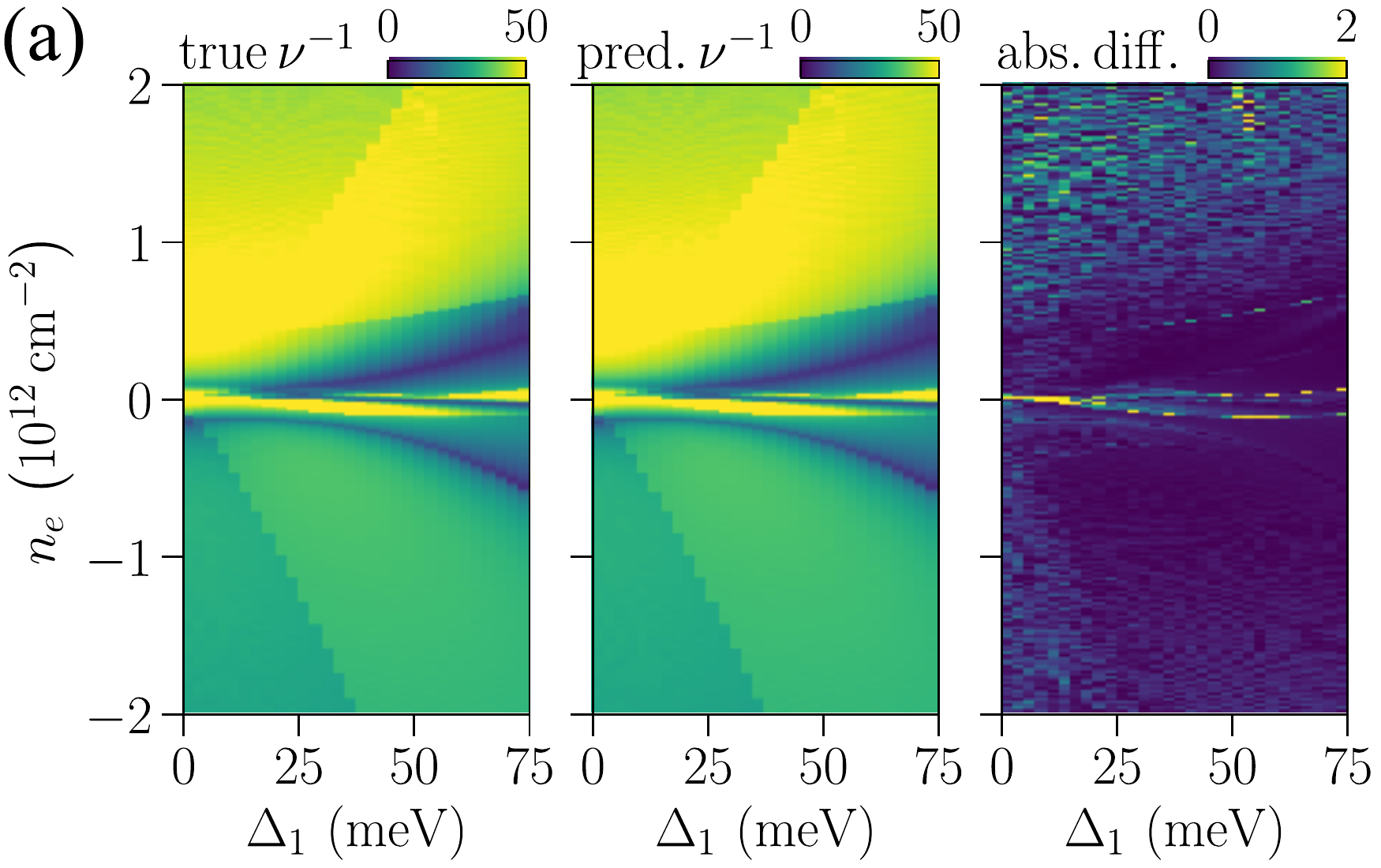}~%
	\includegraphics[width=0.5\linewidth]{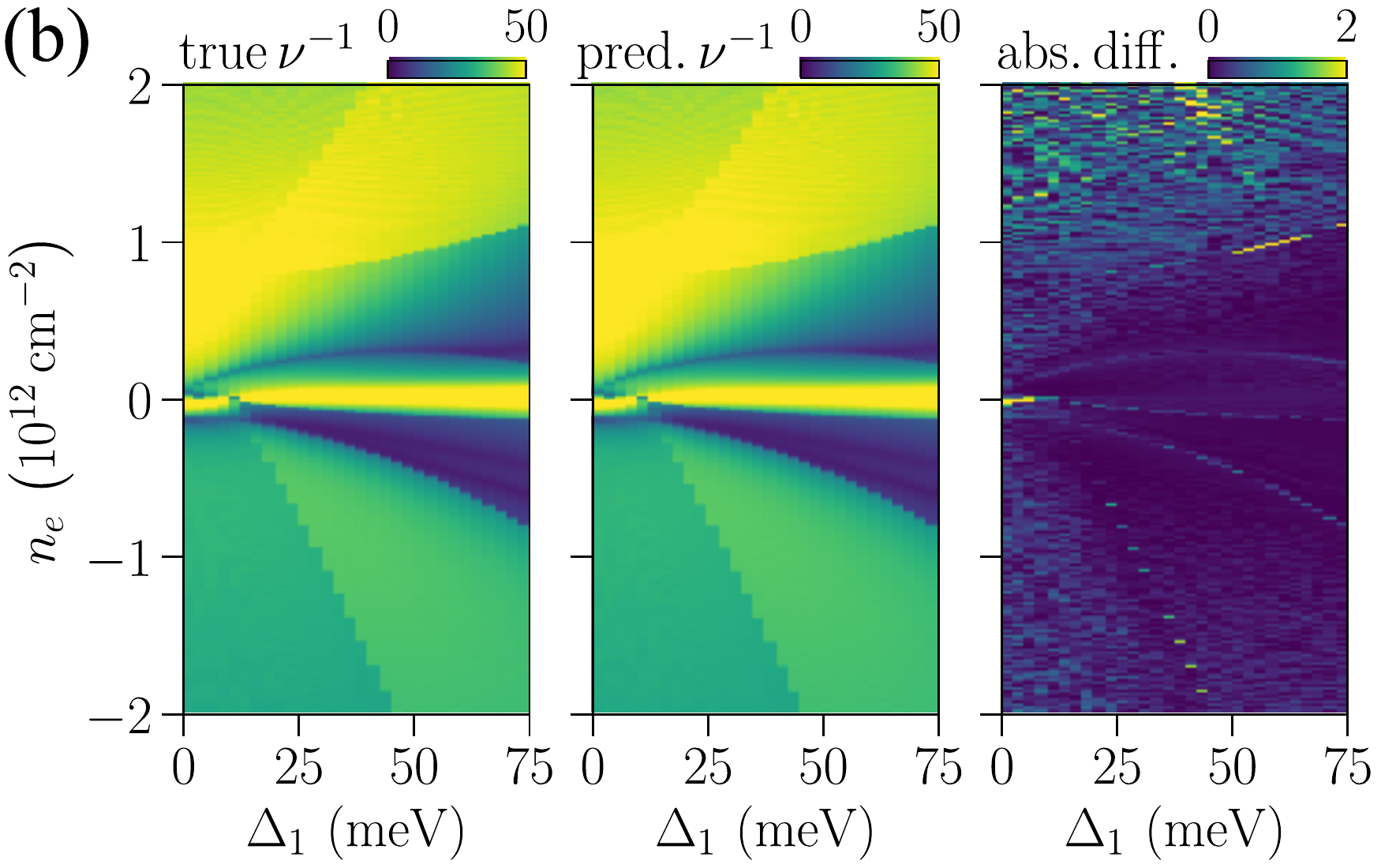}
	\caption{Results on inverting simulator outputs, i.e.~predicting the band-structure parameters that gave rise to a given plot. Each panel shows (left) simulated $\nu^{-1}$ as a function of $\Delta_1$ and $n$, for the specified `true' band parameters; (center) simulated $\nu^{-1}$ for the values of the band parameters inferred by our method; (right) absolute difference between simulations and inferred data. The tables at the top compare the true and predicted values of parameters.
	}
	\label{fig:inverse-sim}
\end{figure*}

\begin{table}[b]
	\centering
	\caption{Quantitative performance on recovering band-structure parameters from simulated data.}
	\label{tab:sim-inv-results}
	\begin{tabular}{p{0.16\columnwidth}@{~~~}>{\raggedright\arraybackslash}p{0.24\columnwidth}@{~~~}>{\raggedright\arraybackslash}p{0.24\columnwidth}@{~~~}>{\raggedright\arraybackslash}p{0.24\columnwidth}}
	\hline\hline
	Parameter & Mean absolute error (meV) & Median abs. error (meV) & Median rel. error \\
	\hline
	$\gamma_2$ & $0.12$ & $0.10$ & $0.0156$ \\
	$\gamma_3$ & $1.54$ & $1.30$ & $0.0512$ \\
	$\gamma_4$ & $0.59$ & $0.45$ & $0.0240$ \\
	$\Delta_2$ & $0.04$ & $0.03$ & $0.0135$ \\
	\hline\hline
	\end{tabular}
\end{table}

\subsection{Inverse problem: determining parameters}
\label{sec:app-inverse}
We next consider the inverse problem, of finding the band-structure parameters corresponding to a set of measurements.
We first evaluate our method on simulated data; this allows validating the proposed approach quantitatively, by measuring how accurately it recovers the parameters that were input to the simulator.
Specifically, we select 100 simulator-generated plots at random from the held-out test set to use as input; each plot shows how DOS $\nu$ varies with $\Delta_1$ and $n$, and we aim to find the corresponding band-structure parameters.

To find the parameters $\bm\gamma$, we follow the procedure described in Section~\ref{sec:method-inverse}, using the same DNN as discussed in Section~\ref{sec:app-fwd}.
As the inputs are simulated data, the DOS itself is directly available, hence we choose the metric $d$ in (\ref{eq:inv-min}) to be the mean squared difference between the input and predicted $\nu$.

Table~\ref{tab:sim-inv-results} shows quantitative results from our approach, calculated over 100 plots from the test set.
For each band-structure parameter, we give the mean and median absolute error, and also the median relative error after first subtracting the midpoint of the corresponding range in Table~\ref{ParTab}.
We see that both absolute values of error and also its relative value are small, with the greatest relative error being $5\%$ for the parameter $\gamma_3$, and even smaller for the remaining parameters. Such a confidence interval in determining tight binding parameters is much smaller compared to the error bars typically available in the literature. For instance, Ref.~\cite{zibrov18} determines $\gamma_3$ parameter with the precision of $17\%$. Further discussion on quantitative performance of the model using parity plots is provided in Appendix~\ref{sec:parity-plot}.
Qualitative illustrations of our results for two particular points in the parameter space are given in Figure~\ref{fig:inverse-sim}. We show the plot provided as input to our method, and the result of running the simulator on the predicted parameter values. Here we show high-resolution momentum grid images for both the input and the prediction, although the DNN operated on low-resolution momentum grid images.
We see that the residual error is very small, and is less concentrated in the vicinity of DOS discontinuities or singularities, compared to Fig.~\ref{fig:forward} that benchmarked the forward approach.

\begin{figure*}[t]
	\includegraphics[width=0.80\linewidth]{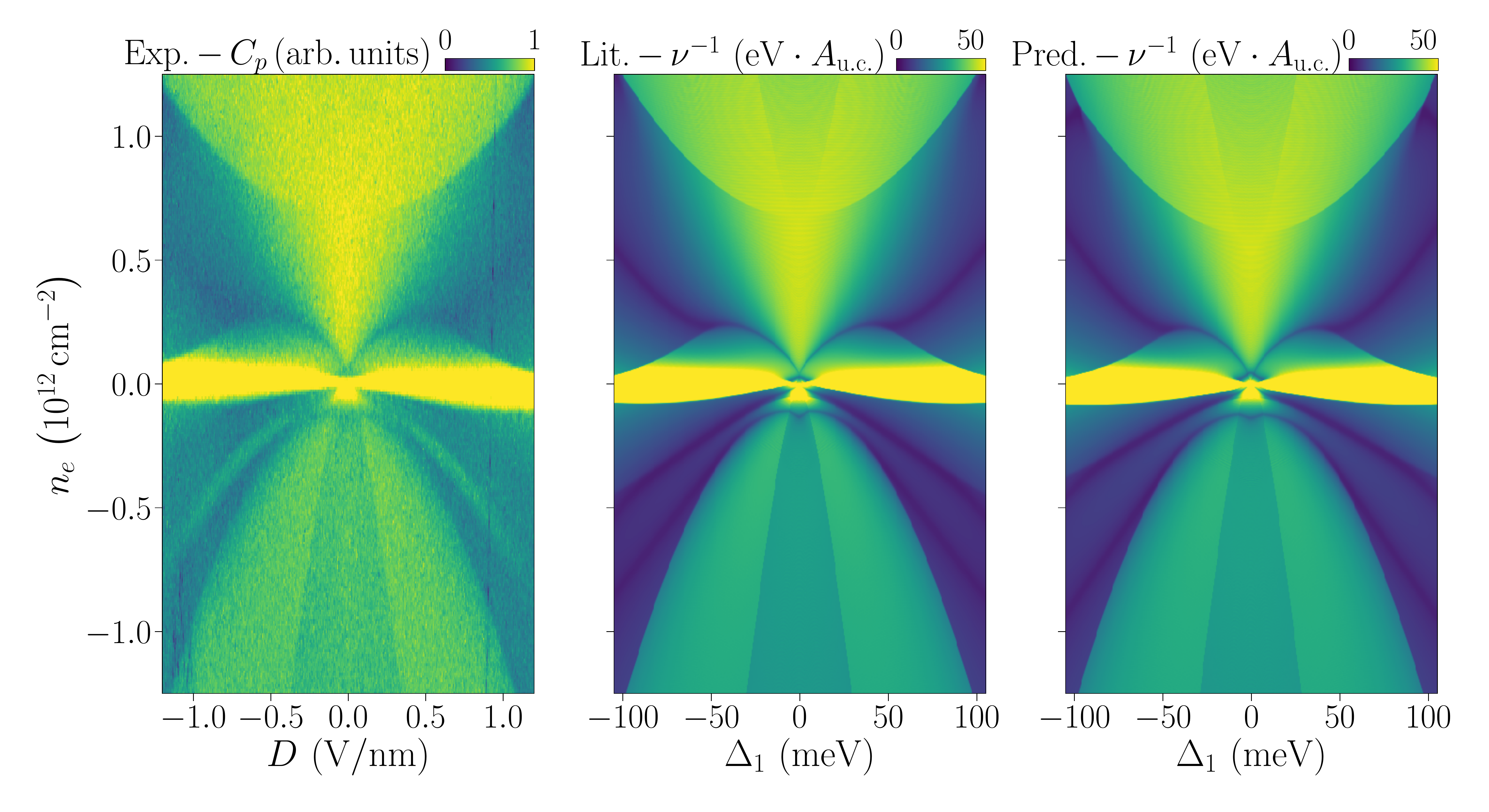}
	\caption{Inferring band-structure parameters from experimental data. The left plot shows experimental $C_p$ measurement as a function of displacement field $D$ and charge density $n$; the middle and right plots show the inverse of DOS $\nu^{-1}$ as a function of $\Delta_1$ and $n$ for parameters $\gamma_2=-21$, $\gamma_3=-290$, $\gamma_4=141$, $\Delta_2=3.5$ and $\gamma_2=-15.5$, $\gamma_3=-312.9$, $\gamma_4=132.3$, $\Delta_2=3.0$ respectively. The middle figure is the best match to the experimental results cited in the literature \cite{zibrov18}, while the right one is the prediction of the DNN. The middle and right plots were obtained using simulator with higher resolution of $\mu$ ($\Delta\mu=0.002$), hexagonal grid with maximum momentum $ka=0.2$, $6.75\times10^6$ grid points, larger range of $\Delta_1$ and with a step of $0.5$.
	}
	\label{fig:inverse-exp}
\end{figure*}

\subsection{Application to experimental data}
\label{sec:app-expt}

Finally, we apply our method to the high quality experimental dataset of penetration capacitance measurements for $ABA$ graphene used in Ref.~\cite{zibrov18}. Note, that Ref.~\cite{zibrov18} utilized additional datasets, namely, the density of states in presence of magnetic field (Landau level regime) in order to assist in determination of tight binding parameters and to infer the associated error bars. In contrast, here we use solely zero magnetic field penetration field capacitance data.
Our goal is to solve a similar inverse problem to the one in Section~\ref{sec:app-inverse}, i.e.~predicting the band-structure parameters for this system, but now given only noisy measurements of $C_p$.

Using the mean squared difference between predicted and experimental $C_p$ for the metric $d$ in Eq.~(\ref{eq:inv-min}) yields poor results in this setting.
This is because features in the simulated and experimental plots (e.g.~jumps in DOS) do not perfectly align for any choice of parameters, and also because there is a small residual difference in the absolute values of $C_p$ between simulated and experimental data even when the plots are optimally aligned. 
These discrepancies can be at least partially attributed to multiple physical mechanisms that are not incorporated in our model. In particular, we take account of the screening only phenomenologically by using the third order expansion~(\ref{Eq:screen}), whereas the use of self-consistent Hartree-Fock screening would result in a more complex function. In addition, we ignore potential renormalization of the band structure parameters by the strong applied electric fields, that could make all band structure parameters dependent on $n$ and $\Delta_1$. Finally, disorder and strain effects are not incorporated in our model, yet these may lead to inhomogeneous broadening of the features.

To mitigate this issue, we adopt techniques from computer vision that are used to align dissimilar images; these are typically employed in tasks such as template matching~\cite{haralick92book,lewis95template}, medical image registration~\cite{studholme96registration,klein10elastix}, and alignment of satellite imagery~\cite{brown92registration}.
Firstly, in order to match prominent features in the plot such as edges regardless of the absolute values of $C_p$, we match the derivative of $C_p$ with respect to $D$ and $n$ instead of $C_p$ itself.
Secondly, instead of the mean squared error, we use the negative zero-normalised cross-correlation \cite{lewis95template}, which is invariant to changes in mean and standard deviation.
We therefore set
\begin{multline}
	\label{eq:zncc} 
	d(C^*_p ,\, C_p) =
	- \frac{
		\partial_{D} C^*_p - m^*_{D}
	}{
		\sqrt{\left\langle( \partial_{D} C^*_p - m^*_{D})^2 \right\rangle}
	}
	\cdot
	\frac{
		\partial_{D} C_p - m_{D}
	}{
		\sqrt{\left\langle( \partial_{D} C_p - m_{D})^2 \right\rangle}
	}
	\\ 
-\frac{
		\partial_n C^*_p - m^*_n
	}{
		\sqrt{\left\langle( \partial_n C^*_p - m^*_n)^2 \right\rangle}
	}
	\cdot
	\frac{
		\partial_n C_p - m_n
	}{
		\sqrt{\left\langle( \partial_n C_p - m_n)^2 \right\rangle}
	}
\end{multline}
in Eq.~(\ref{eq:inv-min}), where $\left\langle \, \cdot \, \right\rangle$ is the mean over all $(D ,\, n)$, $m^*_{D} = \left\langle \partial_{D} C^*_p \right\rangle$, $m^*_n = \left\langle \partial_n C^*_p \right\rangle$, $m_{D} = \left\langle \partial_{D} C_p \right\rangle$, $m_n = \left\langle \partial_n C_p \right\rangle$, and the derivatives of the experimental measurements are approximated using finite differences.
Lastly, in order to mitigate against local minima, we average the objective~(\ref{eq:zncc}) over multiple scales, i.e.~subsampling the grid $Q$ by factors $2^i$ for $i \in \{1 \ldots 5\}$, and correspondingly smoothing and downsampling $C_p$ and $C_p^*$ to form scale-space pyramids \cite{anderson84pyramids}.

Solving the resulting minimization problem as described in Section~\ref{sec:method-inverse} recovers the following parameters:
\begin{eqnarray}\nonumber
\gamma_2 &= &-15.5 \,\mathrm{meV},
\quad
\gamma_3 = -312.9 \,\mathrm{meV},
	\\
\gamma_4 &=& 132.3 \,\mathrm{meV},
\quad
\Delta_2 = 3.0 \,\mathrm{meV}.
\end{eqnarray}
These are broadly consistent with the estimates obtained from manual fitting in Ref.~\cite{zibrov18}.
More precisely, our estimates of $\gamma_2$, $\gamma_3$ and $\gamma_4$ lie within the error bars of \cite{zibrov18} (although $\gamma_2$ is at the upper limit), while $\Delta_2$ is slightly outside (we predict 3.0, compared with their confidence interval $[3.3 ,\, 3.7]$). Notably, the obtained $\gamma_2$ is actually very close to the value obtained for ABC trilayer graphene~\cite{zhou2021half}. 

In Figure~\ref{fig:inverse-exp}, we visualize the experimental data, and the output from the simulator when run with the above parameters at high resolution.
The simulated plot shows a good match to the experimental one, with similar features (e.g.~discontinuities) appearing at similar locations.
We also show the result using the parameters of Ref.~\cite{zibrov18}, these are again qualitatively very similar.
From visual inspection the separation between the two Van Hove singularities in the hole region appears to be wider for predicted parameters than for the one from Ref.~\cite{zibrov18}. This trend is consistent with the experimental results. For all other features the difference between the two models is visually indistinguishable. Therefore, the newly obtained parameters are a better fit to the experiment data.

\begin{table}[b]
\centering
	\caption{
		Quantitative results of sensitivity analysis on experimental data. For each parameter, we record the values to which it must be decreased/increased to cause the distance function $d( C_p^* ,\, C_p )$ to increase by $1\%$.
		We specify both absolute and relative values; a large percentage indicates low sensitivity to that parameter/direction, as a large change is required to affect the quality of fit.
	}
	\label{tab:sensitivity}
	\begin{tabular}{lrrrrr}
		\hline\hline
	Parameter & \multicolumn{2}{c}{Decreasing} & \multirow{2}{*}{Best fit (meV)} & \multicolumn{2}{c}{Increasing} \\
	\cmidrule(lr){2-3} \cmidrule(lr){5-6}
	~ & meV & rel.diff. & ~ & meV & rel.diff. \\
	\midrule
	\hline
	$\gamma_2$ & $-17.25$ & $-11.2\%$ & $-15.51$ & $-14.45$ & $+6.8\%$ \\
	$\gamma_3$ & $-343.5$ & $-9.8\%$ & $-312.9$ & $-296.8$ & $+5.1\%$ \\
	$\gamma_4$ & $111.1$ & $-16.1\%$ & $132.3$ & $153.2$ & $+15.8\%$ \\
	$\Delta_2$ & $2.72$ & $-9.5\%$ & $3.00$ & $3.17$ & $+5.5\%$ \\
	\hline\hline
	\end{tabular}
\end{table}

We also measure the sensitivity of the experimental fit with respect to each of the band structure parameters.
Specifically, we increase and decrease each of the parameters $\gamma_2$, $\gamma_3$, $\gamma_4$ and $\Delta_2$ separately, until the distance $d( C_p^* ,\, C_p )$ between predicted and experimental $C_p$ values increases by more than $1\%$ from the `best fit' value.
This lets us measure quantitatively how large a change in each parameter is required to cause the same reduction in the quality of fit.
Results from this analysis are given in Table~\ref{tab:sensitivity}.
We see that the fit is least sensitive to $\gamma_4$, with changes of around $16\%$ in either direction required to cause a change of $1\%$ in the distance $d$.
Conversely, increasing $\gamma_3$ by just $5.8\%$ leads to a $1\%$ change in $d$; $\Delta_2$ and $\gamma_2$ are similarly sensitive to increases.

\section{Discussion}
\label{sec:discussion}
To summarize, we proposed a deep neural network based approach to determining band structure parameters of two-dimensional materials. Our framework consists of two steps, that rely on the existence of a method to simulate the experimentally accessible data. In the first step we train a deep neural network to obtain a more efficient replacement for the data simulator. In the second step, we extract band structure parameters by minimizing the difference between the experimental dataset and the data simulated by the neural network. To illustrate the application of our framework, we focused on a specific material --- trilayer graphene --- and performed band structure parameter extraction using the experimental data on the penetration field capacitance that effectively probes the density of states. Our procedure resulted in the precise determination of band structure parameters that are close to those determined manually.

In contrast to manual fitting, the approach proposed in our work is more automated, thus involving less human effort. In addition, it is capable of providing more precise values of band structure parameters, and of estimating associated error bars.  We expect that our framework can be easily applied with minimal modifications to the penetration field capacitance data in other two dimensional materials. Moreover, it can be generalized to other experimental probes, such as penetration field capacitance in presence of magnetic field, transport, scanning tunneling microscopy, and other probes that are sensitive to the band structure details.  Application to different experimental quantities requires replacement of the simulator function in our framework, which would be straightforward. A more intricate step consists in understanding the relation between the simulated quantities and experimental measurements. For instance, in the case of penetration field capacitance, the matching of experimental data additionally required incorporating non-trivial screening of electric field, the presence of parasitic capacitance and geometric capacitances of the gates. These steps require physical insight into the details of the experimental measurements.

Finally, a conceptually novel generalization direction may include incorporating interaction effects into our framework. For instance, the recent observation of Stoner transitions in twisted bilayer graphene \cite{zondiner2020cascade,wu2021chern} or chirally stacked $ABC$ trilayer graphene \cite{zhou2021half}, calls for unambiguous identification of interaction parameters alongside the band structure details (that may be in turn quite significantly renormalized due to interactions). We expect that such an extension of our framework may be feasible, provided one is able to construct an efficient simulator of the density of states that incorporates the interactions.  This would constitute an important step towards an experimental extraction of the complete Hamiltonian governing electronic degrees of freedom, thus bringing two dimensional materials on par with quantum simulators that use cold atoms or other platforms to synthetically engineer and verify a desired Hamiltonian.

\section*{Acknowledgments}
A.F.Y. acknowledges primary support from the Department of Energy under award DE-SC0020043, and additional support from the Gordon and Betty Moore Foundation under award GBMF9471 for group operations.

\appendix

\begin{figure*}[t]
\begin{overpic}[width=\linewidth,abs,unit=1mm]{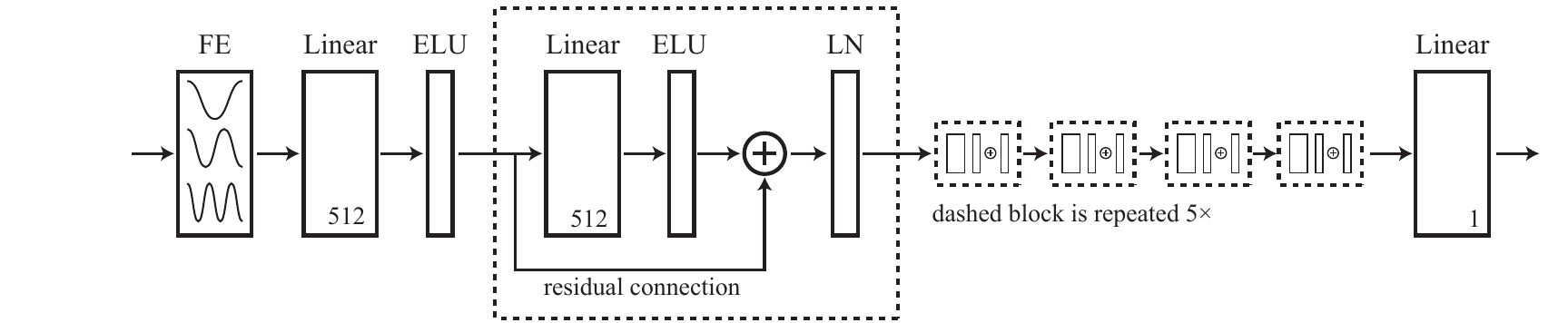}
\put(1,21){$\gamma_{2,3,4}, \Delta_2$}
\put(3,17){$\eta, \Delta_1, n$}
\put(176.5,18.75){$\nu$}
\end{overpic}
\caption{Architecture of the deep neural network $F_\omega$. The DNN takes band structure parameters $(\gamma_2,\, \gamma_3,\, \gamma_4,\, \Delta_2)$, inverse parasitic capacitance $\eta$, interlayer asymmetry $\Delta_1$ and charge density $n$ as input. It processes them with a Fourier embedding layer (FE), linear layer with 512 outputs, and ELU nonlinearity, followed by five blocks with residual connections consisting of linear, ELU, and layer-norm (LN) layers. The resulting features are mapped via a linear projection to the density of states $\nu$.}
\label{fig:architecture}
\end{figure*}

\section{Additional model and training details}
\label{sec:architecture}
	The architecture of our deep neural network $F_\omega$ is shown in Fig.~\ref{fig:architecture}.
	In total, the DNN has $1.35 \times 10^6$ weights (i.e.~trainable parameters).
	We train the model using Adam~\cite{kingma15adam} with a fixed learning rate of $10^{-3}$, and the standard momentum parameters $\beta_1=0.9$, $\beta_2=0.999$ and $\epsilon=10^{-7}$.
	The batch size is set to 512.
	Convergence is determined according to the loss on the held-out validation set -- we stop training once the validation loss has failed to decrease for 10 consecutive epochs, and retain the checkpoint with minimum validation loss.
	We use a single Nvidia RTX~2080~Ti GPU for training, operating at 32-bit precision.

\begin{figure}[t]
	\includegraphics[width=0.90\linewidth]{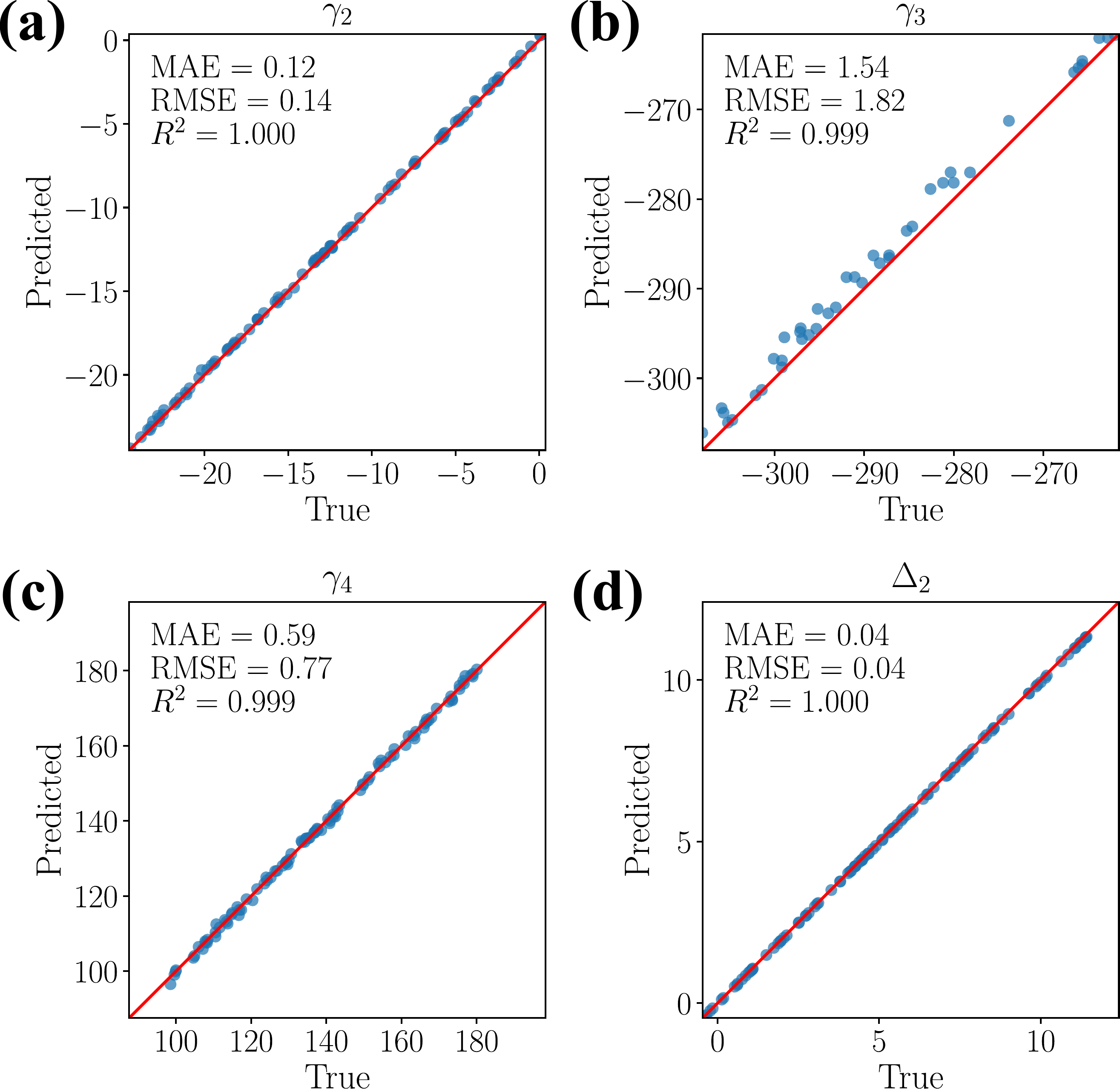}
	\caption{Parity plots of the band structure parameters from simulated data for the inverse problem using the test set of 100 plots. The analyzed data is the same as in Table~\ref{tab:sim-inv-results}. MAE (mean absolute error),  RMSE (root mean square error) and $R^2$ metrics are also shown. Equality line ($y=x$) is shown in red.
	}
	\label{fig:parity-plots}
\end{figure}

\section{Additional quantitative results on the inverse problem}
\label{sec:parity-plot}
To gain further insights on the performance of the model for the inverse problem, Fig.~\ref{fig:parity-plots} shows parity plots of the band structure parameters $\gamma_2$, $\gamma_3$, $\gamma_4$ and $\Delta_2$ for the same test set of 100 plots which was used for obtaining the metrics in Table~\ref{tab:sim-inv-results}. As noted in the discussion of Table~\ref{tab:sim-inv-results} the largest deviation is observed for parameter $\gamma_3$, and Fig.~\ref{fig:parity-plots}~(b) shows that the prediction of the model is always higher than the true values. This systematic error hints at a direction for further improving the model accuracy, and would be an interesting target for future investigation.

\end{document}